\documentstyle[12pt]{article}

\def\be{\begin{eqnarray}}
\def\ee{\end{eqnarray}}
\def\nn{\nonumber}

\begin{document}

\hfill{hepth/9903087}\\
\phantom.\hfill ITEP/TH-12/99

\bigskip

\bigskip

\centerline{\it A.Morozov}

\bigskip

\centerline{117259,  ITEP, Moscow,  Russia}

\bigskip

\centerline{\it Talk given at the Workshop
"Integrability: the Seiberg-Witten and Whitham equations"}
\centerline{Edinburgh, September 1998}

\bigskip

\bigskip

\centerline{ABSTRACT}

\bigskip

Review of the theory of effective actions and the
hypothetical origins of integrability
in Seiberg-Witten theory.

\bigskip

\bigskip

\bigskip

This Conference is devoted to remarkable links between
the recently discovered Seiberg-Witten ansatz \cite{SW}
for the low-energy effective action of $N=2$ supersymmetric
Yang-Mills theories in 4 and 5 space-time dimensions
and the old theory of classical (0+1)-dimensional integrable systems,
based on the notions of Lax representations and spectral
Riemann surfaces. The basic question about this whole subject
is why at all integrability properties should arise in the
study of high-dimensional models, and it remains obscure and
even non-addressed in most publications. Still, in \cite{GKMMM}
integrability has been deliberately searched for -- and found --
as the hidden structure behind the Seiberg-Witten ansatz.
It is the goal of this brief presentation to collect the
implicit arguments in favor of the {\it a priori} existence of a
hidden integrable structure.
These arguments are non-convincing and somewhat non-constructive,
still they deserve being kept in mind in the study of Seiberg-Witten
theory and its predecessors (like matrix models) and possible
generalizations.

\bigskip

{\bf 1. Why integrability?}

The question about the origins of
integrability in Seiberg-Witten theory is in fact a part of the
broader question: What is the role of integrability in physics?
Different people would give different answers to this question,
and  most of them would choose between the following options:

- Integrable models provide funny examples.

- These exactly solvable examples can be (and actually are) used
as the starting points for perturbative expansions describing
approximately the really relevant physical models.

- Whatever is solvable is integrable, i.e. if anybody succeeded
in providing exact solution to some physical model, there should
be a (hidden) integrable structure in this model, responsible
for the very possibility to solve it. This {\it a posteriory}
integrability principle, while certainly true, is not very
interesting: we would prefer to know about the existence of
integrable structure in advance.

- Integrability theory is rather a branch of mathematics
(group theory), and can be of little physical significance.

What we are going to advocate in the present notes is a less
obvious {\it hypothesis}:
Integrability is the general property of effective actions.

However, in order to understand this hypothesis it is important
to have an adequately adjusted (and broad enough)
notion of integrability.

\bigskip

{\bf 2. What is integrability? \cite{UFN2,UFN3}}

Effective actions are functional integrals like
\be
{\cal Z}\{t|\phi_0\} \equiv \int_{\phi_0}
D\phi\ e^{S(t|\phi)}
\label{effact}
\ee
These quantities depend on two types of variables:
on the shape of the ``bare action'' $S(\phi)$ and
on the boundary conditions $\phi_0$ for integration
variables (fields) $\phi$. In ``physical language''
these are respectively dependencies on the ``coupling constants'' $t$
and on the choice of ``vacuum'' $\phi_0$.

Obviously, being a result of integration, an (exponentiated)
effective action ${\cal Z}$ exhibits peculiar features,
unfamiliar from the studies of other classes of objects
in quantum field theory and mathematical physics.
Identification of such features should be the key to
characterization of the new class of special functions,
which will be used to represent the answers to any questions
in non-perturbative quantum field (string) theory.

Among already discovered, the most important are
the two peculiarities \cite{UFN3}:
(i) covariance of ${\cal Z}$ under variation of the coupling constants
(the shape of $S(\phi)$), reflecting invariance of an integral
under a change of integration variables (fields) $\phi$,
and
(ii) certain relation between the dependencies of ${\cal Z}$ on
the shape $S(\phi)$ and on the choice of the vacuum (boundary
conditions) $\phi_0$, reflecting dependence of vacuum states on
the shape of the action.

The feature (i) turns to imply that effective actions exhibit a
(hidden) integrable structure and hypothetically
{\it effective actions are (always?) the (generalized)
$\tau$-functions of integrable hierarchies}.

So far, we have a restricted set of examples where this hypothesis
is explicitly formulated and checked:

- Ordinary matrix models. Here one can straightforwardly study the
$t$-dependence of matrix integrals and find the links to the simplest
KP/Toda-like  $\tau$-functions.

- Character formulas and  Chern-Simons theory. Here the dependence
on boundary conditions can be investigated and again the links
with KP/Toda $\tau$-functions are easily established.
This result can be less expected than the previous one, still
it is true.

- Generalized Kontsevich Model (GKM). The GKM theory allows to
study the interplay between the $t$- and $\phi_0$-dependencies.

- Seiberg-Witten conjecture and Donaldson N=4 SUSY 4d Yang-Mills
model (the topological 4d Yang-Mills theory). In this context the
quasiclassical (Whitham) $\tau$-functions enter the game.

All these examples are distinguished by the property that relevant
$\tau$-functions belong to the well studied KP/Toda family (i.e.
are associated with the simplest Kac-Moody algebra $\hat{U(1)}_{k=1}$).
New examples can be discovered after further advances in the theory
of geometrical quantization when the notion of $\tau$-functions for
other groups becomes more familiar.

\bigskip

{\bf 3. Ordinary matrix model \cite{GMMMO,MM,UFN3}.}

In this case the role of the functional integral is played
by an ordinary integral over $N\times N$ Hermitean matrices $\phi$:
\be
Z\{t\} = \int d^{N^2}\ \phi \exp \left(\sum_{k=0}^\infty
t_k {\rm Tr} \phi^k\right)
\ee

Invariance of the integral under arbitrary change of
integration variable $\phi \rightarrow f(\phi)$ leads to
covariance of $Z\{t\}$ under the change of its arguments:
the coupling constants $t_k$.
Namely \cite{MM}, if $\delta\phi = \epsilon \phi^{n+1}$,
then the action $S(\phi) = {\rm Tr} \sum_k t_k\phi^k$ changes by
$\delta S = {\rm Tr} \epsilon \sum_k kt_k\phi^{k+n}$,
and -- if $\epsilon$ is a number, not a matrix  --
this variation can be reinterpreted as the result
of the change of $t$-variables: $\delta t_{k+n} = \epsilon kt_k$.
This implies the relation  like $Z\{t\} \approx Z\{t + \delta t\}$.
Such relation would imply that $Z\{t\}$
is actually independent of $t$.
In fact, one should also take into account the change of the measure
$d\phi$ under the variation of $\phi$, and exact statement,
expressing covariance of $Z\{t\}$, is:
\be
L_n Z = 0 \ \ \ {\rm for}\ n\geq -1, \nn \\
L_n = \sum_{k=0}^\infty kt_k\frac{\partial}{\partial t_{k+n}} +
\sum_{a+b=n} \frac{\partial^2}{\partial t_a\partial t_b}
\ee
This covariance still implies that $Z\{t\}$ is essentially independent
of its arguments $t_k$, but exact statement is that the dependence
exists, but is of a very peculiar form:
\be
Z\{t\} = \tau_{KP} \{t|G_0\},
\ee
i.e. $Z\{t\}$ is a KP  $\tau$-function at some special point
$G_0$    of the universal Grassmannian (particular group element of
$\hat{U(1)}_{k=1}$). This    statement can    be
checked by explicit evaluation of the matrix integral \cite{GMMMO}
in the formalism of orthogonal polynomials.

Thus, dependence of effective action on the coupling constants
can be represented in terms of the $\tau$-functions.

\bigskip

{\bf 4. Generalized $\tau$-functions and dependencies on the
boundary conditions.}

Generalized $\tau$-function \cite{gentau}
is a generating function for all the
matrix elements of a given group element $G$ in a given representation
$R$:
\be
\tau_R(t|G)   =  \langle 0|  e^{t^aJ^a}  G |0\rangle_R
\ee
Here $t^a$ are parameters of the generating function and
$J^a$ are  generators of the Lie algebra under consideration.

As already mentioned, the conventional KP/Toda-like   $\tau$ functions
are associated with the free fermion realization of the simplest
affine algebra  $\hat{U(1)}_{k=1}$:
$J(z) = \tilde\psi\psi(z)$, $S = \int_{\cal C} \tilde\psi\bar\partial
\psi$. Here ${\cal C}$ is a Riemann surface (complex curve),
which can be  used instead  of the group element
$G = \exp \sum_{m,n} A_{mn}\tilde\psi_m\psi_n$ in the definition
of $\tau$ for affine (1-loop) algebras. Also, in this case it is
convenient to parametrize the set of $t$-variables by a single
function $\bar A$:
\be
\tau_{KP}(\bar A|{\cal C}) = \langle \exp \left(
\int_{\cal C} \bar A J\right) \rangle_{\cal C}
\ee
The standard form of KP tau-function is a particular case of this
one, associated with $\bar A = \bar\partial\varphi$,
$\varphi = \sum_{k=0}^\infty {t_k}{(z-z_0)^{-(k+1)}}$.

The same KP $\tau$-function can be alternatively represented
in terms of a topological {\it three}-dimensional field theory:
abelian Chern-Simons model \cite{M} (this is a simple example of
the ``AdS/CFT-correspondence'' \cite{AdS/CFT}):
\be
\tau_{KP}(\bar A|{\cal C}) = \int_{{\cal A}_{\partial M = {\cal C}}
= \bar A} {\cal DA} \exp S_{CS},
\ee
where
\be
S_{CS} = \int_M {\cal A} d{\cal A} +
\oint_{\partial M} A\bar A
\ee
and $M$ is the ``filled Riemann surface ${\cal C}$''
-- the 3d manifold with the boundary ${\cal C}$.

This identity illustrates the important hypothesis: the boundary
conditions dependence of the functional integral
(that of the Chern-Simons theory in this particular example)
can be also (like that on the coupling
constants) represented in terms of the $\tau$-functions.
It may be not so surprising, because the dependencies on
coupling constants and boundary conditions are in fact
deeply interrelated.

\bigskip

{\bf 5. Generalized Kontsevich Model \cite{GKM,t+T,UFN3}.}

This interrelation can be illustrated with the example of
Generalized Kontsevich Model (GKM), defined as the matrix
integral over $n\times n$ Hermitean matrices $X$,
\be
{\cal Z}_{GKM} \{L|V_{p+1}\} = C\int d^{n^2}X\
e^{{\rm tr}(-V_{p+1}(X) + XL)}
\label{GKMdef}
\ee
Here $L$ is a $n\times n$ Hermitean matrix and $V_{p+1}(x)$
is a polynomial of degree $p+1$, its derivative
$W_p(x) = V'_{p+1}(x)$ is a polynomial of degree $p$.
In GKM theory one introduces two sets of ``time-variables''
to parametrize the dependencies on $L$ and on the shape of
$V_{p+1}(x)$:
\be
T_k \equiv \frac{1}{k}{\rm Tr} \Lambda^{-k}, \  \  \
\tilde T_k \equiv \frac{1}{k}{\rm Tr} \tilde\Lambda^{-k}, \  \  \
L = W_p(\Lambda) = \tilde\Lambda^p
\label{GKMtimes}
\ee
and \cite{Kritimes}
\be
t_k = \frac{p}{k(p-k)}{\rm res}_{\mu} W_p^{1-k/p}(\mu)d\mu
\label{Kritimes}
\ee
In (\ref{GKMdef})
$C$ is a prefactor used to cancel the quasiclassical contribution
to the integral around the saddle point $X=\Lambda$,
\be
C  = \exp\left({\rm tr} V_{p+1}(\Lambda) - {\rm tr} \Lambda
V'_{p+1}(\Lambda)\right)
{\det}^{1/2} \partial\partial V_{p+1}(\Lambda)
\ee

The central result of the GKM theory is that \cite{t+T}
\be
{\cal Z}_{GKM}\{L|V_{p+1}\} =
e^{-{\cal F}_p(\tilde T_k|t_k)}\tau_p(\tilde T_k+ t_k)
\label{t+T}
\ee
where $\tau_p$ is a $\tau$-function of the ``$p$-reduced
KP hierarchy'', and the shape of $\tau_p$ as a function of
time-variables (i.e. the associated group element of $\hat {U(1)}$)
depends only on degree $p$ and not on the shape of the polynomial
$W_p(x)$.
The exponential factor in (\ref{t+T})  contains
\be
{\cal F}_p(\tilde T_k|t_k) = \frac{1}{2} \sum_{i,j} A_{ij}(t)
(\tilde T_i + t_i)(\tilde T_j + t_j), \nn \\
A_{ij} = {\rm res}_\infty\ W^{i/p}(\lambda)dW_+^{j/p}(\lambda)
\label{GKMpreexp}
\ee
which is a ``quasiclassical (Whitham) $\tau$-function''.

At the first glance this GKM example does not seem very close to
(\ref{effact}): neither of the two types of variables in (\ref{GKMdef}) --
the shape of $V_{p+1}$ and the matrix $L$ -- is obviously interpreted
as a boundary condition or vacuum.
Instead the whole expression (\ref{GKMdef}) looks very much like
a (matrix) Fourier transform, and the interrelation like
(\ref{t+T}) between the $L$- and $V_{p+1}$-dependencies, which puts
them essentially on equal footing, is not too surprising:
usually the dependence of Fourier transform (${\cal Z}$) on its
argument ($L$) contains entire information about the shape
of transformed function ($e^{-V_{p+1}(x)}$).
However, the GKM example can be not so irrelevant
for our consideration. As shown in \cite{RGWh}, the GKM
partition function (\ref{GKMdef}) can be a natural member
(the oversimplified case, associated with the genus-zero
spectral curve) of the family of the
Seiberg-Witten prepotentials, which are clearly the quantities
of the type (\ref{effact}).

\bigskip

{\bf 6. The low energy effective actions.}

The low   energy effective actions  describe the effective
dynamics of light modes, arising after all the heavy modes
are integrated away.
In the case  of the Seiberg-Witten
theory the light modes are abelian supermultiplets
(their lowest components parametrize the valleys in the
potential, i.e. the  moduli   space of vacua) and the
Chern-Simons
degree of freedom (not a field!) $K = \int d^3x {\rm Tr}
(AdA + \frac{2}{3}A^3)$, peculiar for Yang-Mills theories
in four dimensions with the property that
(non-perturbative) effective potential
is always a periodic function of $K$. It remains an
unresolved  problem to derive the low-energy effective action for
the fluctuations along the valleys in interaction with the
dynamics in $K$-direction -- which should be just a (0+1)-dimensional
problem    -- directly from the non-perturbative $N=2$ SUSY
Yang-Mills theory.
Still one can attempt to guess  what the transition to the low-energy
effective actions should mean from the point of view of the
$\tau$-functions.

Normally, in higher-dimensional field theories the functional
integrals depend on additional parameter: the  normalization
point  $\mu$, which is the IR cutoff  in  the  integration over
fluctuations with different momenta. Effective action  with
a given $\mu$  describes  the  effective dynamics of excitations
with wavelenghts  exceeding $\mu^{-1}$.
The   low-energy effective
action arises  in the limit   $\mu \rightarrow 0$,
when only finite  number of excitations  (the zero-modes of
the massless  fields) remain relevant.
Of course, different  theories  can possess  the same low-energy
action  (e.g. in all   the theories without massless
particles there are no degrees of freedom,
surviving in the low-energy approximation, and the low-energy
effective action is  just zero for all of them): these actions
are pertinent for  universality classes   rather than for
particular  field theory models. As    usual     in the
study of universality classes      it is  instructive  to look
for  the simplest representative  of the given class.
The matrix integrals, which are             (0+0)-dimensional
models are  such simplest representatives           of the
class, which also contains the $2d$ topological sigma-models
interacting with $2d$ gravity.   Similarly, the Seiberg-Witten
conjecture can be interpreted so that $N=2$ SUSY Yang-Mills models
belong to universality class, of        which the simplest
representatives are the (0+1)-dimensional integrable  systems.

The natural hypothesis (suggested  in \cite{GKMMM}) about  interpretation
of the   low-energy      effective actions in terms of
the $\tau$-functions states that:

- The  low-energy effective  actions  are quasiclassical (Whitham)
$\tau$-functions.

- The  proper     coordinates (the flat structure) on the     moduli
space of vacua are provided   by the adiabatic invariants
$\left( a_C = \oint_C \vec p d\vec q\right)$.

- The time-variables,    associated with the low-energy correlators
(i.e. renormalized coupling constants) are Whitham times
(the     deformations      of   symplectic       structure).

The Seiberg-Witten theory provides us with two kinds of pieces of
evidence in support of this hypothesis:

- Dynamics of branes seems      to  imply the      shapes     of
the spectral        curves   and their period matrices
\cite{Wbra,MaMM,Kap}.

- The correlation   functions  in Donaldson theory seem to be
indeed described in terms of  the relevant Whitham $tau$-functions
\cite{LNS, RGWh}.

\bigskip

{\bf 7. Dynamics of   5-branes \cite{Wbra,Gobra,MaMM,Kap}.}

The fundamental object associated with the $d=11$ supergravity
-- and thus, presumably, with entire      $M$-theory --
is  a      membrane, i.e.    a     2-brane.  Its dual       --
which should  be equally fundamental -- is  a 5-brane
with a 6-dimensional world  volume. In the first-quantization
approach dynamics  of a 5-brane          is thus described
in terms of        some $6d$ field  theory, which -- as a first
choice -- can be either a non-abelian super-Yang-Mills model or
that of abelian self-dual 2-forms.
If the           topology of the    world    volume is
$R^4\times {\cal C}$ one     gets  a  family of  $4d$ Yang-Mills
models, parametrized by the choice   of vacua, labeled  by
Riemann surfaces ${\cal C}$. The possible           choice
of $ {\cal C}$           is restricted by the equations of
motion of the $6d$ theory.
According to \cite{MaMM}, these equations for $6d$ super-Yang-Mills
model, provide ${\cal C}$ in the    form of the spectral curve,
\be
{\cal C}:\ \ \ \det \left( L(\xi) - \lambda\right) = 0
\label{Lax}
\ee
with the flat coordinates on the  moduli space of such curves
defined as $a_C = \oint_C \lambda$.
If  one further considers the theory of abelian 2-forms on {\it such}
curves, one immediately gets  \cite{Wbra} the Seiberg-Witten
prepotential (the $N=2$   SUSY substitute of      the low-energy
effective action) ${\cal F}(a_A)$, of which the second
derivative        is the      period matrix of the spectral curve
${\cal C}$:  $a_B = {\partial{cal F}}/{\partial  a_A}$,
$T = {\partial    a_B}/{\partial a_A}$.

Eq.(\ref{Lax}) is  in fact the equation of motion for
scalar fields $\Phi_{ij}$ in the $6d$ super-Yang-Mills theories:
\be
D^2 \Phi = {\rm fermionic\ v.e.v.}
\ee
For the   topology $R^9\times T$ with $2d$ torus $T$ (the   {\it bare}
spectral curve) and  the  5-brane wrapped around the torus,
this equation becomes (in appropriate gauge $A = 0$ and
$\bar A_{ij} = {\rm diag}(\bar  a_i)$)
\be
(\bar\partial      + \bar A)(\partial\Phi) =\ {\rm source}
\ee
If there      is no source at the                  r.h.s.,
$\Phi_{ij} = {\rm const}$ -- this is the case
of unbroken $N=4$ supersymmetry in four dimensions.
Breakdown of    supersymmetry (down to $N=2$ as result, say,
of non-trivial boundary conditions imposed along $T$
on some fields) somehow produces non-vanishing source at the r.h.s.
If, for example, the   source is a       $\delta$-function,
$m(1-\delta_{ij})\delta^{(2)}(\xi - \xi_0)$, then
\be
L_{ij} = \partial   \Phi_{ij} \sim
d\xi\left(p_i\delta_{ij} + m(1-\delta_{ij})
\frac{\theta\left(\xi - \xi_0 +
\frac{Im\ \tau}{\pi}(\bar a_i -\bar a_j)\right)}
{\theta(\xi - \xi_0)}\right)
\ee
This is the Lax operator of elliptic   Calogero model.

In the double-scaling (Inosemtsev's) limit     $m\rightarrow   \infty$,
$\tau \rightarrow              i\infty$,
$\Lambda^{2N} = m^{2N}e^{2\pi i \tau}$ the    spectral    equation
\be
\det\left(L(z)    - \hat\lambda\right) = 0
\ee
turns into familiar equation for the Toda-chain system:
\be
z + \frac{1}{z} = W(\lambda), \ \ \
\hat\lambda  = \lambda\frac{dz}{z}
\ee
where $W(\lambda)$ is a polynomial of degree $N$ and $z \sim e^{2\pi i\xi}$.

Thus we see, that the brane dynamics forces  the brane,
wrapped  around the {\it bare} spectral curve (with some SUSY breaking
boundary conditions), to split its  $N$ sheets in such a way
as to form  the   {\it full} spectral curve.

\bigskip

{\bf 8. Correlation functions    for $N=2$ SUSY YM in $4d$.}

The     main observables  in Donaldson theory are  made  from
the scalar fields $\phi$:    the lowest components    of the
supermultiplets.
If ${\cal O}_k = {\rm tr}\  \phi^k$, of interest are the
quantities
\be
C_{k_1\ldots  k_n} =
\langle {\cal O}_{k_1} \ldots {\cal  O}_{k_n}\rangle
\ee

The ordinary anomalies in Yang-Mills theory, like
\be
\partial_\mu J_\mu^5 = \beta {\rm tr} F\tilde F, \nn \\
\theta_{\mu\mu} = \beta {\rm tr}  F^2
\ee
imply that
\be
\Lambda\frac{\partial}{\partial\Lambda} C_{k_1\ldots  k_n} =
\beta C_{2\ k_1 \ldots k_n},
\ee
in particular
\be
\Lambda \frac{\partial}{\partial\Lambda}\langle {\cal O}_k\rangle
= \beta \langle {\cal  O}_2 {\cal O}_k\rangle
\ee

In \cite{LNS} the correlation functions
$\langle {\cal  O}_2 {\cal O}_k \rangle$ were explicitly evaluated
by methods   of Donaldson theory
as functions  of the moduli. In \cite{RGWh} it was   shown    that
these answers are indeed consistent with interpretation in terms
of Whitham $\tau$-functions, which implies that
\be
\langle {\cal O}_{k_1} \ldots {\cal  O}_{k_n}\rangle
= \frac{\partial^n {\cal F}(a,T)}{\partial T_{k_1}\ldots
\partial T_{k_n}}
\ee

The prepotential (Whitham $\tau$-function) ${\cal F}$ and the
Whitham times $T$ are introduced by the       following construction
\cite{Kritimes,Kr,IM2,RGWh}. The differential  $dS =\lambda$ which  is
the eigenvalue         of the Lax      matrix 1-form
$L_{ij}    =        \partial\Phi_{ij}$ has      the property that
\be
\frac{\partial  dS}{\partial  {\rm moduli}} = {\rm    holomorphic\
differential\ on}\ {\cal C}
\ee
If one    allows  ${\cal C}$ to have punctures, then ``holomorphic''
means that the first-order poles are allowed at punctures.
If punctures collide,  the higher order poles       are   allowed.
$dS$ can be now deformed:
\be
dS = \lambda + \sum T_kd\tilde\Omega_k
\ee
where   $d\tilde\Omega_k$  has a     pole of   degree      $k+1$   at
a given point. One   can now include the cycles          going  around
the  punctures into the set of $A$-cycles  and those going between the
punctures into the set  of  $B$-cycles. The enlarged     period
matrix $T_{KL}$ can be used to define the prepotential:
\be
T_{KL}    = T_{LK}  = \frac{\partial^2{\cal F}}{\partial T_K\partial T_L}
\ee
The prepotential itself is given by the sum over the    enlarged
set of       $A$ and $B$ cycles:
\be
{\cal F} = \frac{1}{2}\sum_K   \oint_{A^K}  dS\oint_{B_K} dS
\ee
Presumably
it should satisfy the generalized  WDVV equations \cite{WDVV}.

In the  case of   the Toda chain one can explicitly
evaluate the          period matrix \cite{RGWh}:
\be
\frac{\partial^2{\cal F}}{\partial T_m\partial T_n} =
-\frac{\beta}{2\pi i}\left({\cal H}_{m+1,n+1} +
\frac{\beta}{mn}\frac{\partial{\cal H}_{m+1}}{\partial a^i}
\frac{\partial{\cal H}_{n+1}}{\partial a^j}
\partial^2_{ij}\log\theta_E(\vec 0)\right), \nn \\
{\cal H}_{m+1,n+1}      = -\frac{N}{mn} {\rm res}_\infty
W^{n/N}(\lambda)dW_+^{m/N}(\lambda)
\ee
This    result   is in          agreement with the calculation
of \cite{LNS}.   On the other hand, it is obviously a  direct
generalization of eq.(\ref{GKMpreexp})      in the case of GKM.

\bigskip

{\bf   9. Acknowledgments.}

It is a pleasure to  thank H.Braden and other organizers of the
Workshop.

My work is partly supported by the Russian President's grant
96-15-96939 and RFBR grant 98-02-16575.

\end{document}